# Third-Body Stabilization of Supercritical $CO_2$ in CO Oxidation: Development and Application of a ReaxFF Force Field for the $CO/O/CO_2$ System

Emdadul Haque Chowdhury[1], Masoud Aryanpour[2], Yun Kyung Shin[1], Bladimir Ramos-Alvarado[1], Matthias Ihme[3,4,5], Adri C. T. van Duin[1,a)]

[1]Department of Mechanical Engineering, The Pennsylvania State University, University Park, PA16802, USA.

[2]Independent Researcher

[3]Department of Mechanical Engineering, Stanford University, Stanford CA 94305, USA

[4]Department of Photon Science, SLAC National Accelerator Laboratory, Menlo Park, CA 94025, USA

[5]Department of Energy Science and Engineering, Stanford University, Stanford CA 94305, USA

[a)] Corresponding author: acv13@psu.edu

## Abstract

Supercritical $CO_2$ ($scCO_2$) plays a crucial role as a solvent in separation processes, advanced power cycles, and materials processing. Nonetheless, the atomistic comprehension of how the dense $scCO_2$ matrix influences the fundamental reaction of carbon monoxide (CO) is still insufficiently explored. Experimental studies and molecular dynamics (MD) simulations frequently fail to detect the highly reactive, transient intermediates, such as atomic oxygen (O), that drive these reactions. To address this issue, we have developed a novel ReaxFF reactive force field for the $CO_2$/CO/O system. The force field parameters were calibrated using density functional theory and second-order Møller-Plesset calculations to model $CO_2$ crystal properties, intermolecular interactions, bond dissociation curves, and reaction energy barriers. The force field reproduces the cohesive energy of the $CO_2$ crystal, the pressure characteristics of bulk sc$CO_2$, the equation-of-state behavior over a wide pressure–density range, the pressure dependence of the C–O bond length under compression, and the structural properties of liquid and sc$CO_2$, as documented by experiments, ab-initio MD, and prominent non-reactive models. The force field was subsequently applied to study the CO + O → $CO_2$ reaction. In a dilute environment, the reaction is inefficient as the newly formed $CO_2$ rapidly dissociates due to excess kinetic and potential energy acquired from the exothermic reaction. Conversely, in a dense sc$CO_2$ environment, the surrounding matrix acts as an efficient third body, stabilizing the emerging $CO_2$ product via molecular collisions. Statistical analysis confirms an average excess energy dissipation of 133.9 ± 3.6 kcal/mol over 112.4 ± 17.9 ps. Kinetic energy decomposition reveals that ~92% of the excess kinetic energy is stored in internal (rotational and vibrational) degrees of freedom. This ReaxFF force field establishes a mechanistic foundation for third-body stabilization in dense reactive environments.

## 1. Introduction

High-pressure supercritical fluid (SCF) technologies are becoming important in modern industry, with applications ranging from separation processes, advanced power generation, to sustainable chemical processing.[1,2] A fluid is classified as supercritical when it is maintained at a temperature and pressure above its critical point, leading to a unique set of physical properties.[3] A substance in this state exhibits liquid-like density and solvating capacity, while maintaining gas-like low viscosity and high diffusivity.[1]

Among these, supercritical carbon dioxide (sc$CO_2$) is one of the most commonly utilized SCFs due to the combination of some highly advantageous properties.[4–6] Its mild critical conditions (31.1°C and 73.8 bar) makes it attractive for thermally sensitive materials processing.[7,8] In addition, its non-flammability, non-toxicity, and abundance in high purity earn it a "generally recognized as safe" (GRAS) status[3]. Compared to conventional solvents, it exhibits superior mass transfer and penetration capabilities owing to its low viscosity, high diffusivity, and negligible surface tension.[9,10] Furthermore, its solvent properties can be tuned by temperature and pressure variation, which can alter the reaction kinetics, making it an appealing reaction medium.[11] This renders it a promising candidate for carbon capture and storage (CCS).[12] In addition, its contribution in the energy sector is becoming increasingly important. The advantageous thermophysical properties render it a suitable working fluid for advanced power cycles, facilitating enhanced thermal efficiency and more compact turbomachinery.[7,11] Quantum mechanical (QM) studies indicate that $scCO_2$ functions not only as an inert diluent but also actively influences combustion by stabilizing reactive intermediates and providing alternative reaction pathways.[13]

Given the active role of $scCO_2$ in modifying combustion pathways and stabilizing intermediates, understanding the chemistry of key species such as carbon monoxide (CO) is essential for optimizing the operation of $scCO_2$-based reaction systems. CO is an important pollutant and intermediate, and its accumulation poses both environmental and health hazards.[14,15] In direct-fired $scCO_2$ power cycles and other advanced combustion systems, complete CO oxidation is crucial for operational stability and efficiency.[16] Experiments and kinetic analyses demonstrated that the primary reaction for CO oxidation in hydrocarbon flames is: $CO + OH \rightleftharpoons CO_2 + H$; and the conversion of CO via $CO + O_2 \rightarrow CO_2 + O$ is slow and insignificant in the presence of a large concentration of radicals (OH, H, O, $HO_2$).[14,17,18] In strongly backmixed CO-air systems and in high-temperature oxy-fuel or $scCO_2$ flames, atomic oxygen can persist at super-equilibrium levels, contributing directly to CO oxidation and indirectly to OH regenerating through O + $H_2O$ → 2OH and O + $H_2$ → OH + H.[16,17,19–21] As such, a comprehensive understanding of how O radicals govern CO oxidation in $scCO_2$ flames is indispensable for accurate kinetic modeling.

Numerous experimental studies have been conducted to elucidate the kinetics of CO combustion.[17,22,23] While such investigations yielded significant insights into macroscopic observables (rates, ignition delays, flame speed, burnout times), they failed to illuminate the underlying radical chemistry.[14,17,24] The high reactivity and short-lived O radicals and other intermediates, such as $CO_3$, pose significant challenges for experiments to determine their precise collision dynamics, particularly at high pressures and in $scCO_2$. These constraints necessitate the utilization of computational methods to interpret the experimental observations and elucidate the intricate reaction mechanisms. Researchers extensively leveraged molecular dynamics (MD) simulations to investigate the atomistic details of $scCO_2$ and combustion systems.[25–31] Primarily, those studies employed classical, non-reactive force fields and *ab*-initio MD to examine the thermodynamic and structural characteristics of $scCO_2$, including solubility, induced dipole effects, phase behavior and its mixtures with co-solvents.[25–27,32] Investigations on the impact of $scCO_2$ on ignition delay and combustion characteristics demonstrated that $CO_2$ dilution modifies flame dynamics in $CH_4$ systems.[33] These force fields are useful for understanding physical behavior, including structural dynamics, thermomechanical properties, and failure mechanisms of materials,[34–43] but they rely on predetermined atomic connectivity and cannot capture dynamic bond formation and dissociation events.[12] Ab-initio simulations, on the other hand, are too computationally costly to study solvent-mediated radical stabilization, which requires large system sizes and time scales.[12,44]

The ReaxFF reactive force field was developed to bridge this critical gap.[45–47] By employing a bond-order-based framework, it can track continuous formation and breaking of chemical bonds, rendering it an effective tool to investigate complex chemical reactions. ReaxFF parameters are trained extensively on QM data, allowing it to achieve accuracies comparable to QM methods with substantially reduced computational cost. This capability allows ReaxFF to simulate millions of atoms over nanoseconds, making it effective for investigating chemical reactions in a bulk supercritical environment.[48–53] ReaxFF has been effectively applied to a range of combustion phenomena, including the pyrolysis and high-temperature oxidation of hydrocarbons, such as methane and propene,[45] the oxidation of methanol in supercritical environments,[33] the oxy-fuel combustion and pyrolysis of coal,[54,55] syngas combustion,[56] and oxidation of iron surfaces in sc$CO_2$.[57] For example, oxy-fuel coal combustion in $O_2$/$CO_2$ environments elucidated $CO_2$ formation mechanisms and demonstrated that elevated $CO_2$ concentrations enhance reaction rates while lowering $O_2$ consumption.[58] In addition, oxy-combustion of syngas in sc$CO_2$ demonstrated modified ignition chemistry and underscored the catalytic functions of $CO_2$.[33,56,59]

Furthermore, ReaxFF simulations revealed that solvents do not always serve as inert environments, but can play a crucial role in stabilizing highly energetic intermediates and products via collisional energy transfer and solvent cage effects, e.g., ReaxFF simulations of methanol oxy-combustion in supercritical $H_2O$ (sc$H_2O$) environment indicate that sc$H_2O$ acts as an effective third body to dissipate excess energy and promote stable product formation, and significantly accelerates methanol decomposition.[33] Another ReaxFF study of Fe oxidation in sc$CO_2$ demonstrated temperature-dependent direct participation of sc$CO_2$ in surface reactions.[57] At 600 K, sc$CO_2$ adsorbs on the iron surface, forming a passivating $FeCO_3$ layer. At 1000 K, however, $scCO_2$ decomposes into CO, O, and C adatoms, resulting in FeO and $Fe_3C$ formation. A recent ReaxFF study indicated that supercritical xenon facilitates iodine recombination by transferring excess energy.[60] These examples underscore the ability of supercritical solvents in product stabilization and their active involvement in reaction chemistry. Likewise, the dense $scCO_2$ matrix may significantly aid in stabilizing the nascent high-energy $CO_2$ molecule produced from the exothermic reaction of CO and O. Despite its significance, a ReaxFF analysis elucidating the atomistic aspects of the $CO + O \rightarrow CO_2$ reaction and the influence of bulk $scCO_2$ environment on stable product formation has not been adequately explored. To address the current knowledge gap, this article introduces a novel ReaxFF reactive force field developed for the $CO/O/CO_2$ system. This force field is employed to predict physical properties of $CO_2$, including the equation of state and cohesive energy, as well as to simulate the reaction dynamics of CO + O in both a dilute and a dense $scCO_2$ environment. This comparative analysis aims to clarify the role of the $scCO_2$ solvent as a third body in stabilizing highly exothermic reaction products.

## 2. Methodology

### 2.1. ReaxFF Force Field

ReaxFF is a bond order-based force field that can accurately represent bond formation and breaking events across various materials and chemical systems.[45,47] ReaxFF dynamically assesses connectivity via bond orders derived from updated interatomic distances at each MD iterations, whereas traditional nonreactive force fields depend on fixed, predefined connectivities. ReaxFF is typically parameterized mainly against quantum mechanical (QM) reference data, enabling it to attain near-QM accuracy while functioning at a significantly faster computational speed. This efficiency facilitates the simulation of large systems across prolonged time scales. Moreover, non-reactive force fields typically employ fixed atomic charges that remain constant throughout the MD simulations, but ReaxFF incorporates a polarizable charge equilibration method (EEM or ACKS2)[61] that recalculates atomic charges at each iteration based on the local chemical environment. ReaxFF determines the total system energy that governs atomic forces using the following equation:

$$E_{sys} = E_{bond} + E_{over} + E_{under} + E_{lp} + E_{val} + E_{tor} + E_{vdWaals} + E_{coulomb} \quad (1)$$

This total system energy($E_{sys}$) accounts for bond formation and dissociation ($E_{bond}$), over/under coordination penalties ($E_{over}/E_{under}$), lone-pair energy ($E_{lp}$), valence angle strain ($E_{val}$), and dihedral angle strain ($E_{tor}$), as well as non-bonded van der Waals ($E_{vdWaals}$) and electrostatic interactions ($E_{coulomb}$). For a more detailed description of the ReaxFF potential functions, readers are referred to the supporting information of Chenoweth et al.[45]

ReaxFF bridges the gap between higher-level quantum chemistry and conventional nonreactive force fields by allowing bond breaking and bond formation at much lower computational cost. Its continuous bond-order formalism makes it especially useful for describing reaction pathways, transition-state regions, and energy barriers more realistically than fixed-connectivity force fields.[45,47,62] Indeed, the ability to reproduce QM-derived reaction energy barriers is a core design objective of ReaxFF, and this has been demonstrated across a wide range of chemical systems, including hydrocarbon oxidation and combustion chemistry.[45,56,63] However, ReaxFF remains an empirical model whose accuracy is inherently lower than that of the QM data used for training, such as DFT calculations at the B3LYP/6-311G** level or higher-level ab initio methods such as CCSD(T). Accordingly, ReaxFF is generally expected to reproduce overall energetic and structural trends with reasonable fidelity, rather than matching every reference data point to QM-level precision.

Its predictive capability also depends strongly on the quality and balance of the training set. Therefore, transferability is limited, and a parameter set optimized for one chemical system may not remain reliable for chemically different reaction environments without further refinement.[45,47,62] Regarding the charge equilibration scheme, commonly used approaches such as EEM can show deficiencies for far-from-equilibrium geometries, especially during bond dissociation, and may lead to unphysical long-range charge transfer.[45,47,62] More advanced approaches such as ACKS2 can reduce this issue, although at greater computational expense.[61] ReaxFF also does not account for electronically excited-state effects, which may become relevant under extreme shock or detonation conditions. These limitations are well recognized in the ReaxFF community and are the subject of ongoing methodological development. Accordingly, the present force field should be viewed as a targeted parameterization for the CO/O/$CO_2$ system and sc$CO_2$ combustion conditions studied here, and its predictions are interpreted in that context throughout this work.

### 2.2. Force Field Parameter Training

The ReaxFF force field is system-dependent and necessitates parameter optimization for precise representation. The parameters are typically calibrated using experimental observations or quantum mechanical reference data, which are typically obtained from density functional theory (DFT) calculations. Before conducting extensive MD simulations of CO oxidation in sc$CO_2$, a dedicated ReaxFF force field parameter set for $CO_2$ was developed. All DFT calculations were performed in the Amsterdam Modeling Suite (AMS).[64] Unless otherwise mentioned, all those calculations are done with the B3LYP hybrid exchange-correlation (xc) functional with a TZ2P basis set. We also included second-order Møller-Plesset (MP2) calculations in our training dataset. Some of the MP2 calculations were incorporated from existing literature, while others were performed in GAUSSIAN[65], using 6-311++G** basis sets. The detailed procedures for parameter training are discussed in Section 3. The MD simulations described in Section 3.2 utilized the AMS-ReaxFF solver, whereas those in Section 3.3 employed LAMMPS.[66] Atomistic visualizations were performed in OVITO.[67]

To obtain the optimized ReaxFF parameter set that best reproduces the reference data, the parameters yielding the minimum cumulative normalized error were selected. The cumulative normalized error was computed using the following equation:

$$Error = \sum_{i=1}^{n}\left[\frac{x_{i,QM} - x_{i,ReaxFF}}{\sigma}\right]^2 \quad (2)$$

where σ denotes the weighting factor assigned to each data point according to its relative importance in the optimization, *n* is the total number of data points in the training set, and *x* represents the target property, mainly, the energy values used in this study. The training set comprises nearly 500 data points, the majority of which are presented in Figures 1–6 and Figure S1 of the manuscript. As mentioned, the weighting of each data point is governed by an accuracy target (σ in Equation 2), which reflects the degree of fidelity desired for that particular data point. In general, tighter accuracy targets (e.g., σ = 0.1 kcal/mol) are assigned to data points with small absolute energy values, such as the crystal deformation energies in Figure 1 and the MP2-predicted $CO_2$ dimer interaction energies in Figures 2 and 3, where precise reproduction is critical. Conversely, looser accuracy targets (e.g., σ = 1–5 kcal/mol) are applied to data points with larger energy magnitudes or where strict matching is less critical, such as the $CO_2$ – $CO_2$ unstable dimer formation pathway in Figure 5(a) and the energy-density EOS derived from the Cygan potential. In addition, when multiple data sources describe the same physical quantity, weights are assigned based on the relative reliability of each source. Furthermore, accuracy targets were selectively tightened for data points where preliminary force field predictions showed notable deviations from the reference values, in order to bias the optimization toward better reproduction of those critical regions. It should be noted that the assigned accuracy targets do not guarantee that the final force field will reproduce every data point within the specified tolerance. Since the optimization simultaneously minimizes the cumulative normalized error across the entire training set (Equation 2), the final parameter set represents the best global compromise. As the accuracy target appears squared in the denominator of the error function, data points with smaller σ values contribute more heavily to the total error, and the optimizer therefore prioritizes their reproduction accordingly. The final ReaxFF parameter set can be found in the Supplementary Information (SI).

## 3. Results and Discussion

### 3.1. The Force Field Development

The force field parameter optimization commenced with training against various crystal deformation versus energy equations of state (EOS) data, which included the bulk modulus and stiffness tensor components sourced from the Materials Project.[68] We employed the lattice parameters, bulk modulus (5 GPa), and stiffness tensor (C) of a cubic $CO_2$ crystal containing four molecules, as illustrated in Figure 1(a). The $CO_2$ crystal measures 5.5 Å on each side.

Initially, we subjected the original $CO_2$ crystal to various deformation types[69], including volumetric deformations (uniform expansion and compression) as well as different crystal strains, such as normal and shear strain. The energy related to volumetric deformations was calculated using the Birch-Murnaghan EOS,[70] while a Taylor expansion[71] was applied to characterize the energy associated with various crystal strains.

The Birch-Murnaghan EOS, a model describing the relationship between the volume and energy of a crystal under compression or expansion, was used to fit the energy-volume data for isotropically distorted structures of $CO_2$. Here, the initial structure was uniformly expanded and compressed up to 12% of its original volume. Figure 1(b) depicts the resulting energy vs. volume EOS. The third-order Birch-Murnaghan EOS is expressed as:

$$P(V) = \frac{3\,B_0}{2}\left[\left(\frac{V_0}{V}\right)^{\frac{7}{3}} - \left(\frac{V_0}{V}\right)^{\frac{5}{3}}\right]\left\{1 + \frac{3}{4}(B_0' - 4)\left[\left(\frac{V_0}{V}\right)^{\frac{2}{3}} - 1\right]\right\} \tag{3.1}$$

where $P$ (GPa) is the pressure, $V$ (Å$^3$) is the volume, $V_0$ (Å$^3$) is the reference volume at zero pressure, $B_0$ (GPa) is the bulk modulus at zero pressure, $B_0'$ (dimensionless) is the first pressure derivative of the bulk modulus.

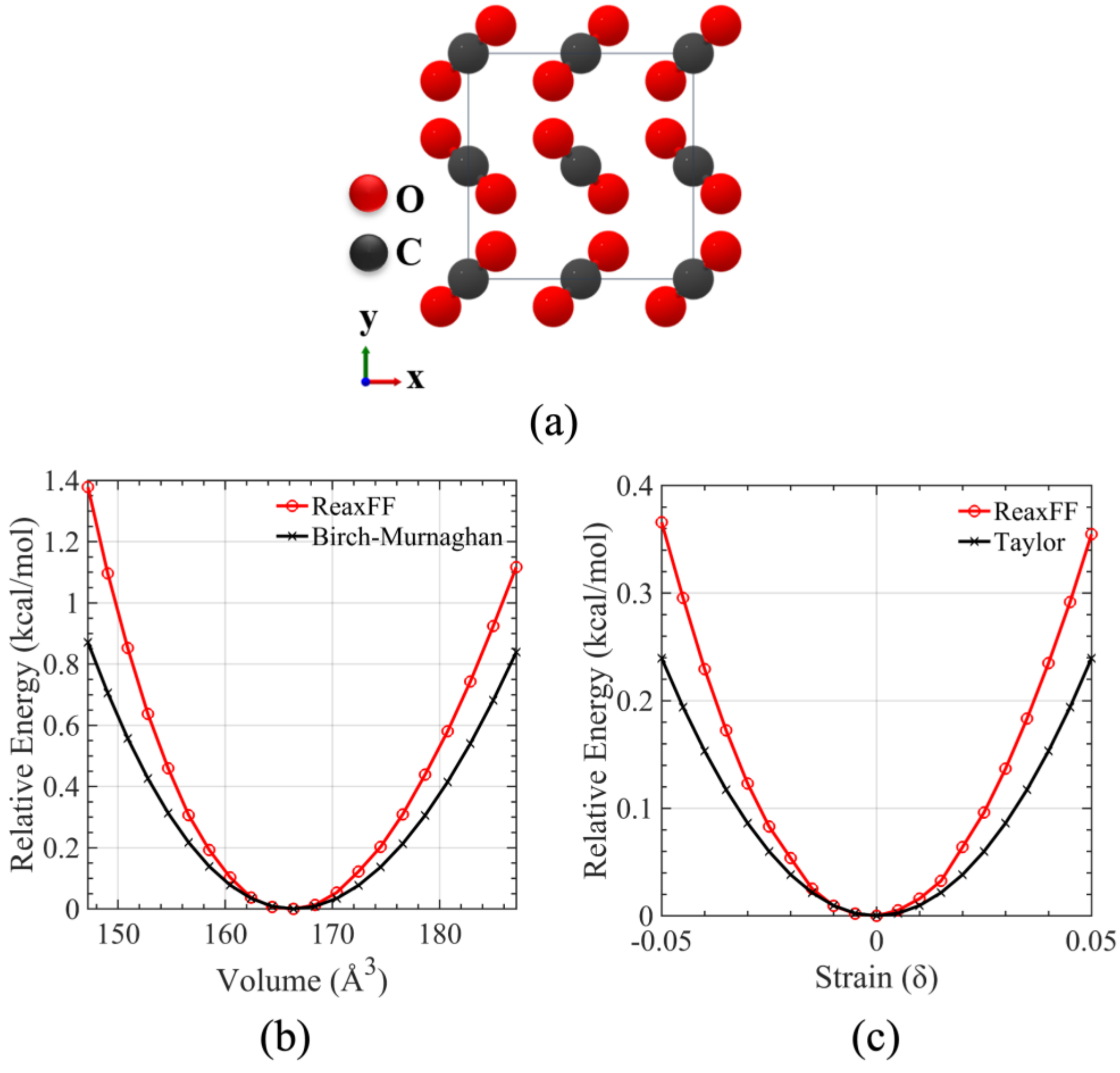

**Figure 1.** Training ReaxFF against $CO_2$ crystal equation of states for bulk modulus and stiffness tensor components. Red and gray spheres represent oxygen and carbon atoms, respectively. (a)

The cubic $CO_2$ crystal considered for this calculation; (b) variation of the relative energy as a function of volumetric strain; and (c) variation of the relative energy as a function of normal strain along the x-axis. The energy of the undistorted $CO_2$ crystal has been taken as the reference for obtaining relative energy values.

The stiffness tensor ($C$) for the considered crystal is given as:[68]

$$C\ (\text{GPa}) = \begin{bmatrix} C_{11} & C_{12} & C_{13} & 0 & 0 & 0 \\ C_{21} & C_{22} & C_{23} & 0 & 0 & 0 \\ C_{31} & C_{32} & C_{33} & 0 & 0 & 0 \\ 0 & 0 & 0 & C_{44} & 0 & 0 \\ 0 & 0 & 0 & 0 & C_{55} & 0 \\ 0 & 0 & 0 & 0 & 0 & C_66 \end{bmatrix} = \begin{bmatrix} 8 & 3 & 3 & 0 & 0 & 0 \\ 3 & 8 & 3 & 0 & 0 & 0 \\ 3 & 3 & 8 & 0 & 0 & 0 \\ 0 & 0 & 0 & 2 & 0 & 0 \\ 0 & 0 & 0 & 0 & 2 & 0 \\ 0 & 0 & 0 & 0 & 0 & 2 \end{bmatrix} \tag{3.2}$$

Here, $C_{11}$, $C_{22}$, and $C_{33}$are the normal stiffness components along the principal axes; $C_{12}$, $C_{13}$, and $C_{23}$ are the coupling stiffness components; and $C_{44}$, $C_{55}$, and $C_{66}$ are the shear stiffness components. These stiffness tensor components represent the material's response to different types of mechanical strains, critical for understanding the material's elastic behavior. Like the Bulk modulus, we applied different types of strain (5%) to the crystal corresponding to each stiffness tensor component, such as normal strain and shear strain. The energy corresponding to each deformed structure was generated using the Taylor series expansion. The Taylor expansion for the internal energy E(V, δ) of a crystal under strain δ up to the second order is given by:

$$\mathrm{E}(\mathrm{V}, \delta) = \mathrm{E}(V_0, 0) + V_0 \left( \sum \tau_i \delta_i + \frac{1}{2} \sum C_{ij} \delta_i \, \delta_j \right) \tag{3.3}$$

where, $\mathrm{E}(V_0, 0)$ (kcal/mol) is the internal energy of the unstrained crystal, $V_0$ (Å$^3$) is the reference volume of the unstrained crystal, $\tau_i$ (GPa) are the components of the stress tensor, $C_{ij}$ (GPa) are the elastic stiffness constants, $\delta_i$ and $\delta_j$ (dimensionless) are the strain components.

Figure 1(c) depicts the resulting strain vs. energy EOS for the $C_{11}$ stiffness tensor components. The *x*-axis in the graphs represents the percentage and direction of applied strain. The EOSs for other stiffness components are presented in the SI, Figure S1. The objective of training the force field parameters in response to the energy changes associated with crystal deformation is to ensure that the force field accurately reflects the elastic properties of $CO_2$. Figure 1 indicates that ReaxFF shows considerable accuracy in reproducing the "U-shaped" energy vs. deformation profiles. Overall, ReaxFF demonstrates good accuracy for less deformed structures, but deviations increase with higher strain. However, it is important to realize that the ultimate accuracy of the force field is contingent upon the entirety of the training set, rather than any one data set. Each component of the training data is designed to collectively guide the parameters in the appropriate direction.

To accurately model intermolecular $CO_2$-$CO_2$ interactions, we trained our force field against $CO_2$ dimer energies for both slid-parallel ($C_{2h}$) dimers (see Figure 2(a)) and cross or T-shaped ($C_{2v}$) dimers (see Figure 3(a)) computed by MP2 calculations, obtained from Ref. [72]. These MP2 energy values were calculated using the 6-311+G(2df) basis set, and basis set superposition errors (BSSE) were corrected by the counterpoise method. It is worth noting that $C_{2h}$ and $C_{2v}$ dimers are reported to be the most stable forms of $CO_2$ dimers.[72] Figure 2 depicts the comparison of ReaxFF and MP2 energy values for the slid-parallel dimers, where $R_1$ is the intermolecular C-C distance in the vertical direction, and $R_2$ is the C-C distance in the horizontal direction. Based on the analysis of Figure 2, the ReaxFF force field demonstrates good performance in predicting the trends of energy-distance profiles. The largest average deviation is observed for dimers with $R_1$ = 3.2 Å (~ 14%), while the smallest average deviation is observed for $R_1$ = 3.4 Å (~ 3%).

Similarly, training data for the $CO_2$ cross or T-shaped dimers, obtained from ref. [72], are presented in Figure 3. Figure 3(a) illustrates the T-shaped dimer, and Figure 3(b) shows the relative energy vs C-C distance (R). ReaxFF is found to mimic the shape of the energy profile considerably well. However, ReaxFF exhibits greater accuracy for distances greater than 4.2 Å, while it slightly underpredicts the potential energy between 3.9 Å and 4.2 Å. The average percentage deviation of ReaxFF for the cross-dimer energy is ~ 10%.

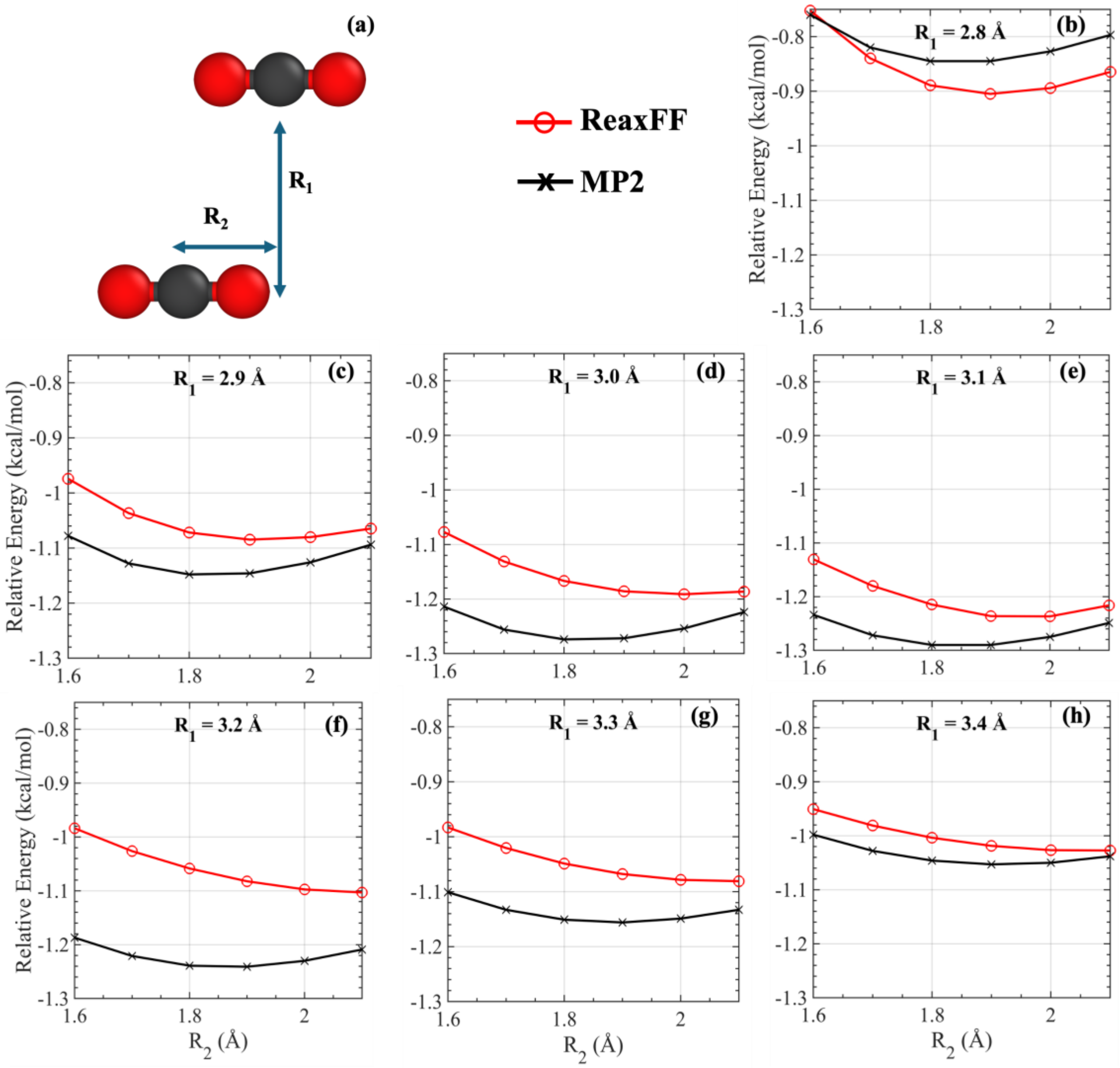

**Figure 2.** Training ReaxFF against $CO_2$ slid-parallel dimer computed from MP2 calculations.[72] (a) Geometry of the slid-parallel dimer considered for this calculation. Red and gray spheres

represent oxygen and carbon atoms, respectively. $R_1$ is the vertical C-C distance, and $R_2$ is the horizontal C-C distance. (b)-(h) Comparison between MP2 and ReaxFF energies. For each plot, the distance $R_2$ is varied by keeping $R_1$ fixed (specified inside the plot). The energy of two isolated $CO_2$ molecules has been taken as the reference for obtaining relative energy values.

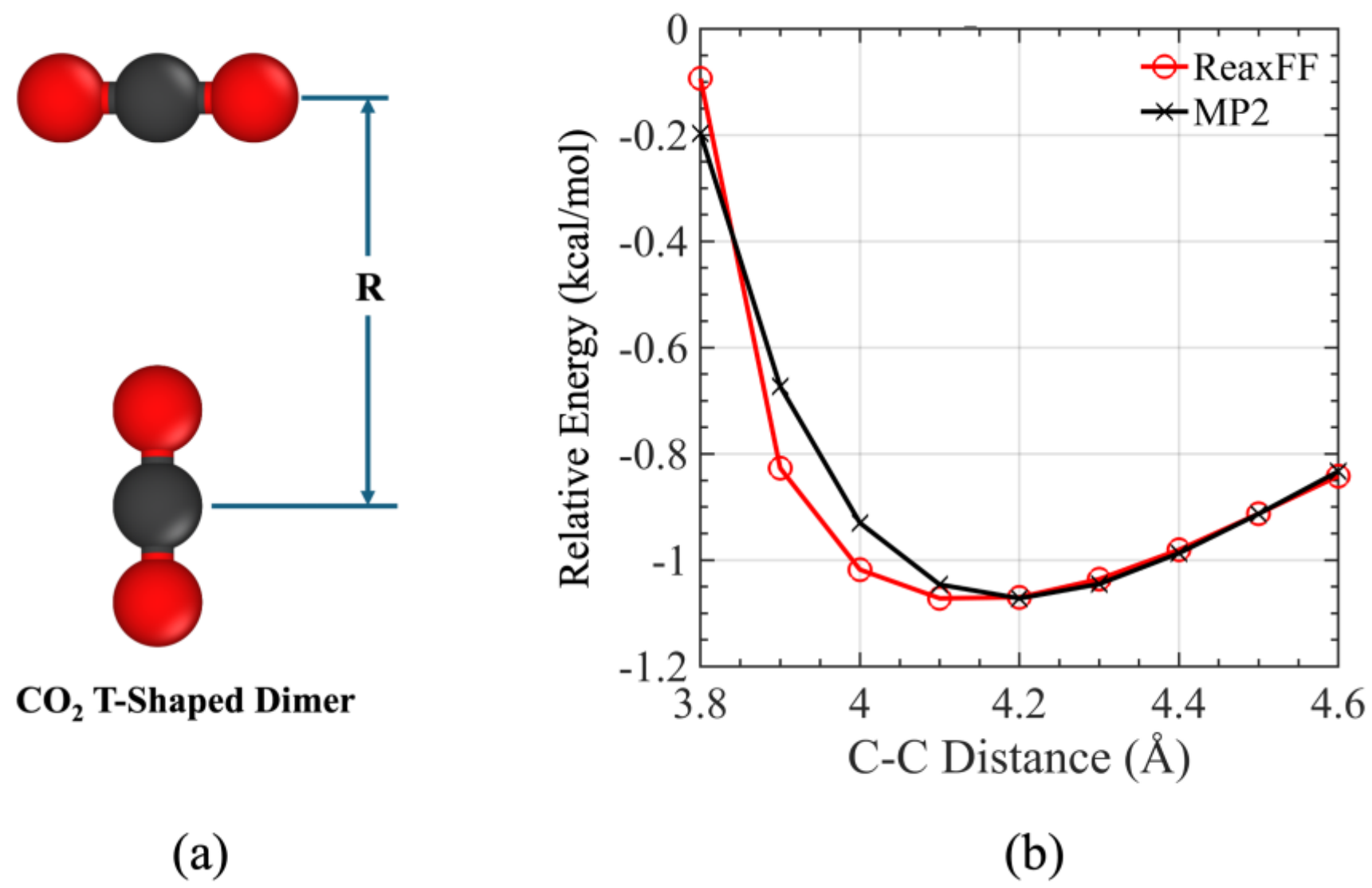

**Figure 3.** Training ReaxFF against the $CO_2$ T-shaped dimer computed from MP2 calculations.[72] (a) Configuration of the $CO_2$ T-shaped dimer. Red and gray spheres represent oxygen and carbon atoms, respectively. The C-C distance is R. (b) Comparison between ReaxFF and MP2 energy vs C-C distance (R). The energy of two isolated $CO_2$ molecules has been taken as the reference for obtaining relative energy values.

Furthermore, to accurately model the pressure and density of $CO_2$, it is crucial to correctly account for long-range intermolecular $CO_2$ interactions. Standard DFT functionals often fail to accurately account for long-range electron correlation effects, responsible for van der Waals interactions.[73,74] This leads to inaccuracies in systems where dispersion forces play a crucial role, such as in molecular dimers. To overcome this issue, more accurate quantum chemical methods, like the MP2 method were used for calculating the $CO_2$ dimer long-range interactions.

Hence, we trained our force field parameters for $CO_2$ dimer interaction energies from 3 Å – 6 Å for both slid-parallel dimers (Figure 4(a)) and cross dimers (Figure 4(b)) using GAUSSIAN, MP2 calculations with 6-311++G** basis sets. In general, ReaxFF slightly underpredicts the stability for parallel dimers, and the accuracy of ReaxFF improves when the C-C distance exceeds 5 Å. Alternatively, ReaxFF shows an excellent agreement in predicting the energy for cross dimers; however, the deviation increases when the C-C distance goes beyond 4 Å. For the parallel dimers, the ReaxFF energy values exhibit an average deviation of ~ 11%, whereas for the cross dimers, the average deviation is ~ 5%.

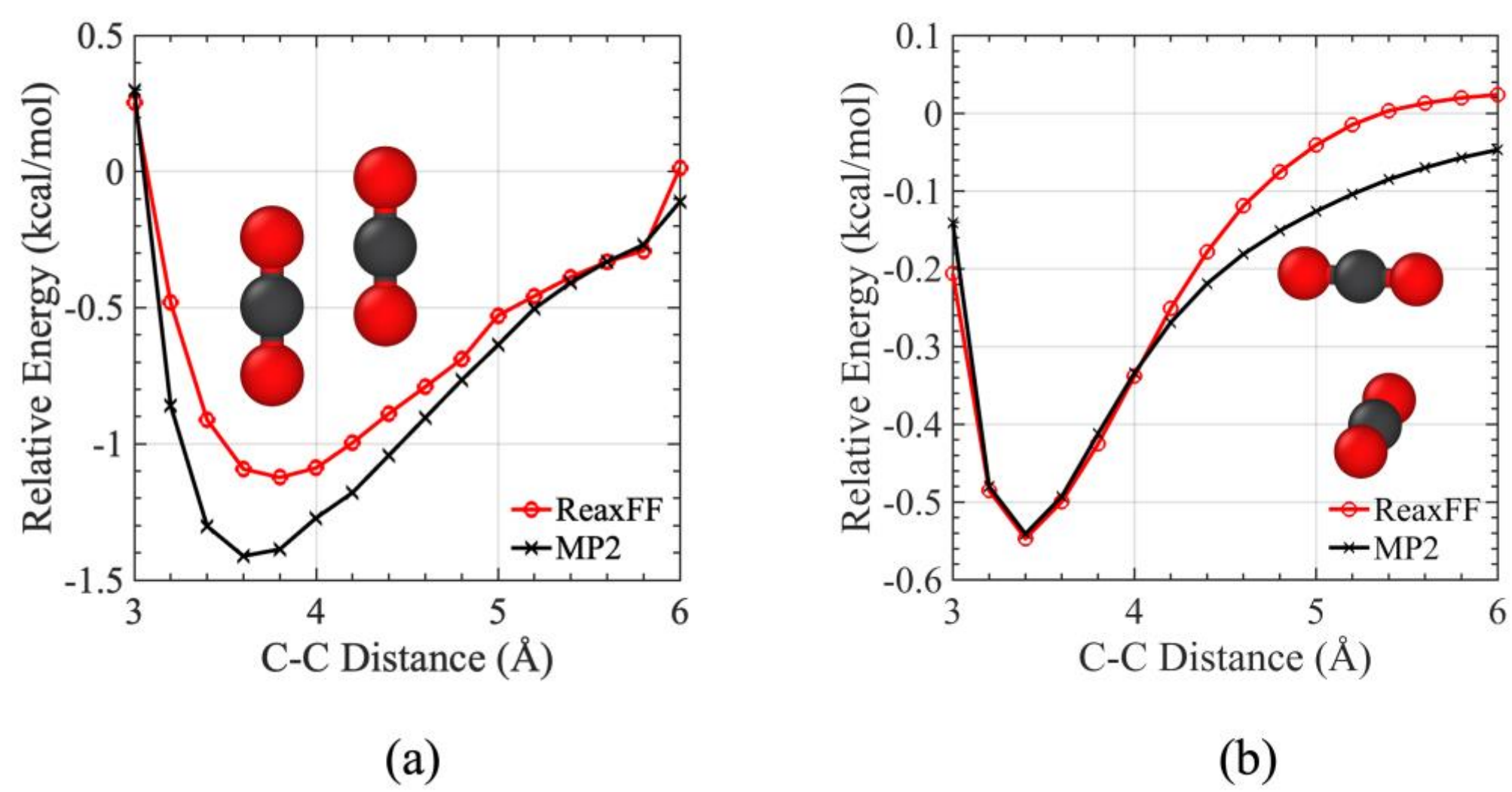

**Figure 4.** Training ReaxFF against $CO_2$ parallel and cross dimer interactions from MP2 calculations. (a) Comparison of ReaxFF and MP2 energies for the parallel dimers. (b) Comparison between ReaxFF and MP2 energy for cross dimers. Red and gray spheres represent oxygen and carbon atoms, respectively. The energy of two isolated $CO_2$ molecules has been taken as the reference for obtaining relative energy values.

To suppress unphysical $CO_2$ dimerization to $C_2O_4$ during MD simulations, it was necessary to train the force field to have a high dimer formation energy barrier and to correctly predict the endothermicity of this reaction. To achieve this, we trained the force field against the DFT-predicted $CO_2$ dimer formation reaction pathway, see Figure 5(a). For this training, the intermolecular C-O distance was varied from 4 Å to 1.2 Å (indicated with a double-headed arrow in Figure 5(a)), and the potential energy of 17 intermediate points was calculated. The DFT calculations were performed using the B3LYP hybrid xc functional with the TZ2P basis set. The DFT estimated dimerization energy barrier was approximately 50 kcal/mol. Figure 5(a) illustrates that ReaxFF traces the energy barrier with sufficient accuracy, indicating that unwanted dimerization is highly improbable. The ReaxFF-predicted energy barrier was approximately 57 kcal/mol. Figure 5(a) also indicates that ReaxFF slightly overpredicts the attraction energy of dimers between 2.6 Å and 3.2 Å compared to DFT. Below 2.6 Å, the ReaxFF energy graph becomes steeper than the DFT graph, indicating a rapid increase in the repulsive energy.

Previous experimental and theoretical investigations indicated that short-lived carbon trioxide ($CO_3$) may form under certain conditions through the interaction of O with $CO_2$.[75–79] Its formation has been experimentally observed under specific conditions, such as during ozone photolysis in liquid $CO_2$[77] and within low-temperature solid $CO_2$ exposed to electron irradiation.[78,79] Despite its brief existence, the production route of $CO_3$ is pertinent to atmospheric phenomena, including $O(^1D)$ quenching and isotope exchange.[78,79] Therefore, accurately capturing the energetics of the $CO_3$ formation reaction was considered important for the accuracy of our ReaxFF force field. Hence, we trained the force field against the $CO_3$ formation energy. We gradually reduced the distance between an O radical and the C atom of a $CO_2$ molecule and used the energy profile obtained from DFT to train ReaxFF. For this DFT calculation, we used the hybrid B3LYP functional with the QZ4P basis set. Figure 5(b) illustrates the trained force field predicting the reaction path for $CO_3$ formation accurately.

To improve the ability of the force field to describe the dissociation behavior of key bonded interactions, we included in the training set the O=O double-bond dissociation curve and the O–O single-bond dissociation curve, shown in Figures 5(c) and 5(d), respectively. The O=O dissociation curve was generated by systematically varying the distance between the two oxygen atoms, while the O–O dissociation curve was obtained by varying the O–O distance in an $H_2O_2$ molecule. The resulting force field reproduces these dissociation profiles with good accuracy. In addition, the C≡O triple-bond dissociation curve of carbon monoxide, obtained from DFT calculations, was also included in the training set. As manifested in Figure 5(e), the trained force field exhibits excellent agreement with the reference DFT data in reproducing the single-well-shaped energy profile. The average deviation of ReaxFF from the DFT data is ~6.4%.

We also trained our force field for the cohesive energy of solid $CO_2$ obtained from existing literature,[80] where authors utilized an incrementally corrected periodic local MP2 (LMP2) scheme with post-MP2 CCSD(T) corrections and basis-set limit extrapolation. Cohesive energy is a fundamental property that indicates the strength of the interactions holding a solid together. It was calculated as the difference between the potential energy of solid $CO_2$ and the sum of the energies of its isolated constituent molecules. This energy is crucial for understanding the stability, binding strength, and various physical properties of materials, like hardness, melting point, mechanical strength, and phase transitions. The ReaxFF predicted $CO_2$ cohesive energy is approximately -7.1 kcal/mol, while the literature value is approximately -7.05 kcal/mol.[80] This good agreement between ReaxFF and literature cohesive energy helps ensure that the developed force field accurately captures the essential intermolecular forces and the overall stability of the $CO_2$ crystal structure.

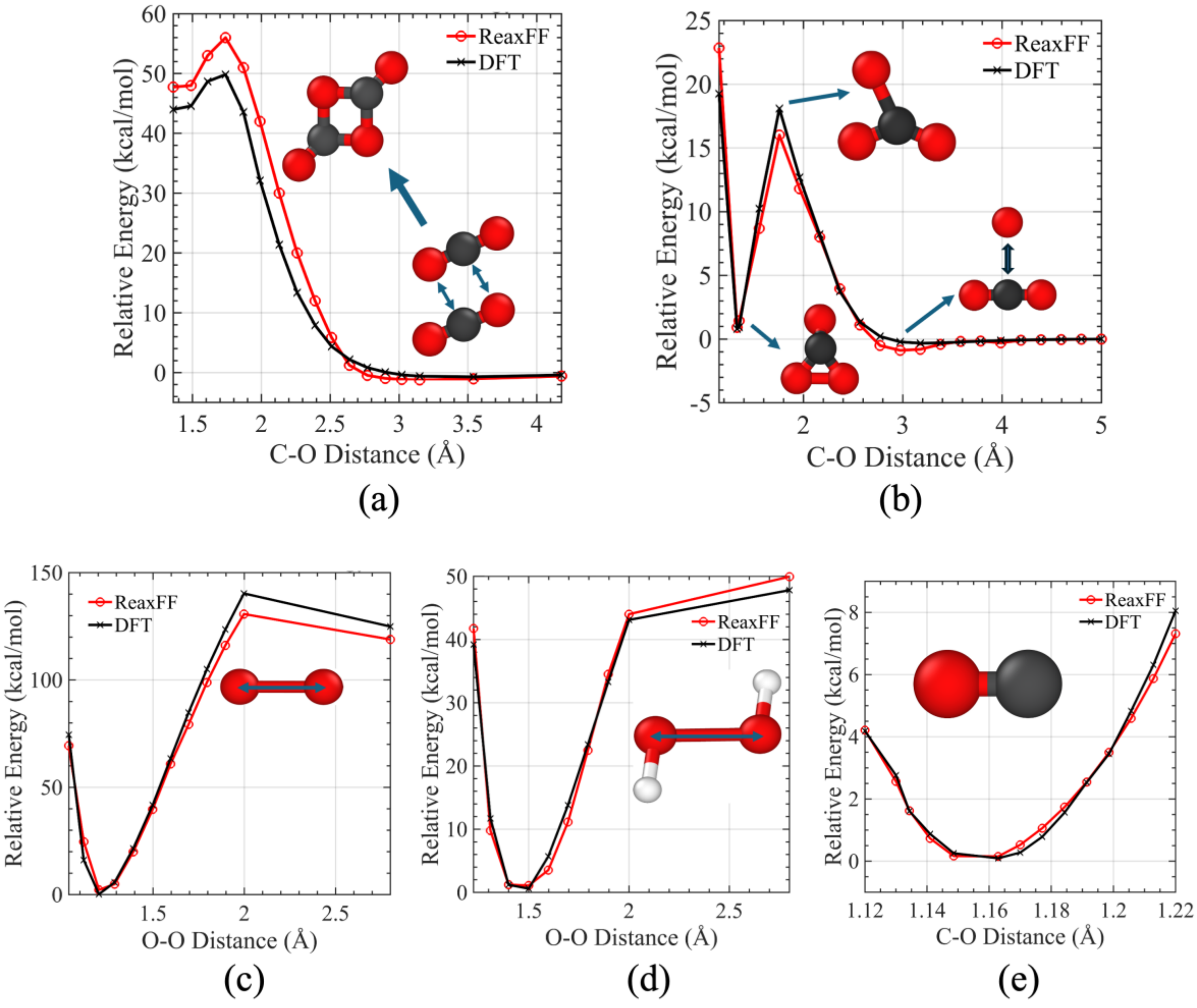

**Figure 5.** (a) Comparison between ReaxFF and DFT for $CO_2$-$CO_2$ dimer formation energy profile. $CO_2$ molecules were brought close to each other by reducing the intermolecular C-O distance, as shown with a double-headed arrow. The energy of two isolated $CO_2$ molecules has been taken as the reference for obtaining relative energy values. (b) Comparison between ReaxFF and DFT for the $CO_3$ formation energy, where a $CO_2$ molecule and O radical were brought close to each other by reducing the intermolecular C-O distance, as shown with a black double-headed arrow. The energy of an isolated $CO_2$ and an O radical has been taken as the reference for obtaining relative energy values. (c) Comparison between ReaxFF and DFT for the O=O double bond dissociation curve. (d) Comparison between ReaxFF and DFT for the O-O single bond dissociation curve. (e)

Comparison between ReaxFF and DFT for the C≡O triple bond dissociation curve. For (c)-(e), the energy of the most stable $O_2$, $H_2O_2$, and CO molecules has been taken as reference energy values, respectively. Red, gray, and white spheres represent oxygen, carbon, and hydrogen atoms, respectively.

To further improve the density and pressure prediction of the force field, we recognized the need to train it for larger $CO_2$ clusters. However, performing QM calculations for a large system is computationally infeasible. Therefore, we chose the Cygan potential[81] that is a validated three-site flexible force field for $scCO_2$ simulations to generate the energy vs density EOS for training ReaxFF parameters. It is worth mentioning that multiple studies have confirmed its accuracy for density and pressure prediction against NIST data.[82–84] Chen et al.[82] ranked it among the most accurate models for liquid-vapor coexistence curves, while Liao et al.[84] reported excellent transport property predictions (self-diffusion, viscosity) across supercritical conditions up to 200 MPa, which justifies its use for generating $scCO_2$ configurations for ReaxFF parameter training. Hence, we conducted NPT simulations at 10 MPa and 316 K ($scCO_2$) with a system of 25 $CO_2$ molecules and another system containing 50 $CO_2$ molecules using the Cygan potential[81] for 3 ns with a 0.25 fs time-step so that the systems reach equilibrium at the supercritical state. We then systematically expanded and compressed the final geometries to generate systems with varying densities from vapor to solid zones (0.14 g/cm$^3$ - 1.8 g/ cm$^3$). This provides short and long-range interactions of $CO_2$ molecules in a larger system. Afterward, we energy minimized each geometry using the conjugate gradient scheme and calculated the energy versus density EOS for those systems using the Cygan potential, and trained ReaxFF parameters against those values. Figures 6(a) and 6(b) show the energy vs. density EOS comparison between ReaxFF and the Cygan potential for the 25 $CO_2$ and 50 $CO_2$ systems, respectively.

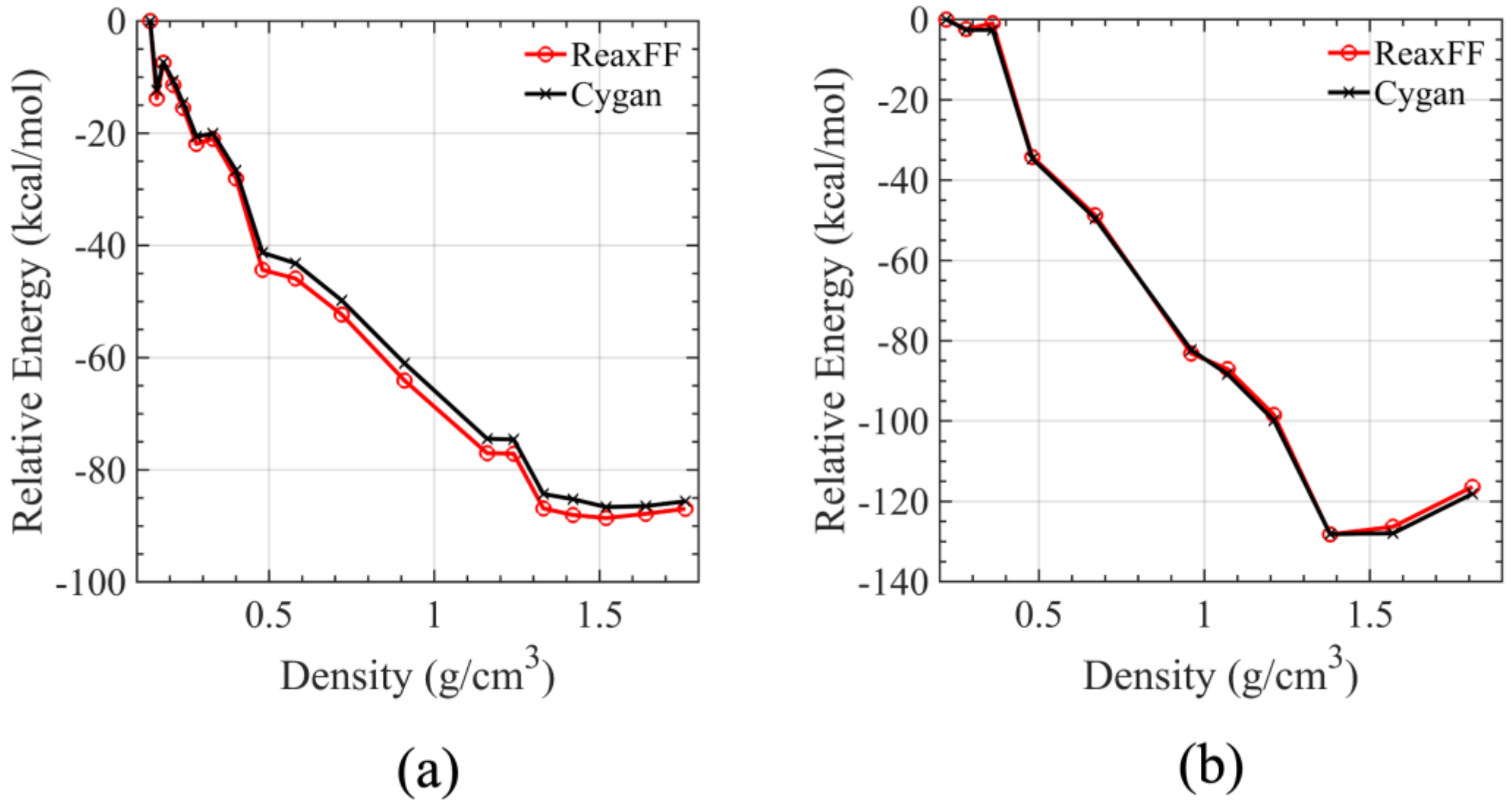

**Figure 6.** Training ReaxFF against energy vs density EOS calculated from Cygan force field for: (a) a system of 25 $CO_2$; (b) a system of 50 $CO_2$.

As expected, as the density increases from vapor to solid, the potential energy decreases continuously due to the stronger intermolecular attraction. The equilibrium density of a $CO_2$ crystal is ~1.52 g/cm$^3$.[68] Hence, further compression beyond this point increases the potential energy values. For the 25 $CO_2$ system, the average deviation of ReaxFF is ~9%, while it is ~3% for the 50 $CO_2$ system.

### 3.2. Testing Force Field Performance

To evaluate the fidelity of our ReaxFF description for predicting the $CO_2$ non-reactive phase diagram, we performed canonical-ensemble (NVT) MD simulations in the supercritical regime (330-440 K). The MD simulations were run for 2 ns with a time step size of 0.25 fs to ensure the systems reached equilibrium. For all simulations, we employed the Berendsen thermostat with a temperature-damping constant of 100 fs. Equilibrium pressures were subsequently calculated by averaging the densities over the last 1 ns of the simulations. For each temperature, two initial configurations with 11,000 $CO_2$ molecules were prepared with densities taken from the National Institute of Standards and Technology (NIST)[85] data along the 20 and 30 MPa isobars. After equilibration at the target temperature, we computed the time-averaged system pressure and compared it directly to the corresponding NIST target (20 or 30 MPa) to quantify deviations. Figure 7(a) illustrates the configuration of supercritical $CO_2$ (11,000 molecules) in the simulation cell. Figure 7(b) depicts the ReaxFF equilibrium pressure versus temperature (330-440 K) for systems initialized at NIST densities on the 20 MPa (blue) and 30 MPa (orange) isobars. Note that the error bars represent the standard deviation of pressure fluctuations inherent to MD simulations, which reflect natural thermodynamic fluctuations in a finite system rather than experimental measurement uncertainty or force field inaccuracy. Overall, ReaxFF predicted $CO_2$ pressure values show good agreement with NIST. Across the investigated temperature range, ReaxFF predicts equilibrium pressures within ~10% of NIST values at 20 MPa and within ~7% at 30 MPa on average. In addition, to further evaluate the equation-of-state behavior of the developed force field, we performed another set of NVT simulations for the same system of 11,000 $CO_2$ molecules at 360 K using several initial densities corresponding to 10-20 MPa according to NIST data. These simulations were run for 1 ns using the same time step and temperature-damping constant as described above. After equilibration, the average system pressure was calculated and directly compared with the corresponding NIST values, as shown in Figure 7(c). The results show good agreement between ReaxFF and NIST over the investigated density range, with an average deviation of approximately 5.62%. To further assess the structural fidelity of the present ReaxFF force field under compression, static energy minimization of the cubic $CO_2$ crystal structure (same as Figure 1(a)) was performed across a pressure range of 0–20 GPa, and the resulting C-O bond lengths are compared against the MP2 reference data of Li *et al.*[86] in Figure 7(d).

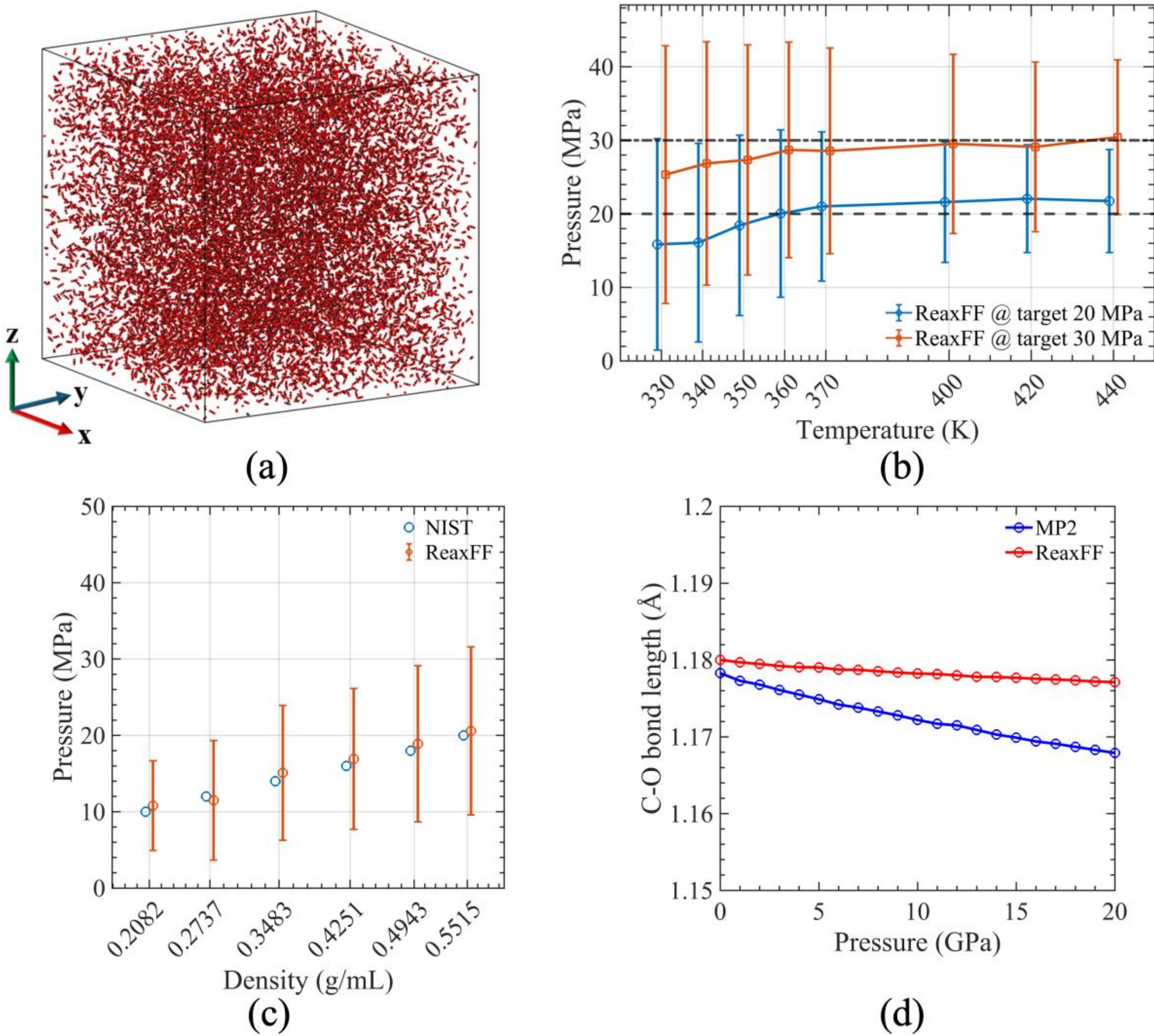

**Figure 7.** (a) Representation of supercritical $CO_2$ (11,000 molecules) in a simulation cell. (b) ReaxFF equilibrium pressure versus temperature for systems initialized at NIST densities on the 20 MPa (blue) and 30 MPa (orange) isobars. Symbols show time-averaged pressure with standard deviation of the pressure fluctuations. Horizontal dashed lines indicate the corresponding NIST targets. (c) ReaxFF-predicted equilibrium pressure versus density for supercritical $CO_2$ at 360 K, compared with corresponding NIST values over the 10-20 MPa range. (d) Pressure dependence of the C–O bond length in cubic CO2-I predicted by the present ReaxFF force field, compared with the MP2 results of Li *et al*.[86] over the 0-20 GPa range.

For this purpose, the cubic $CO_2$ crystal was systematically compressed in LAMMPS using the fix box/relax command at each target pressure, followed by structural optimization with the conjugate gradient minimization scheme, and the average C–O bond length was then extracted from the relaxed crystal structure. As shown in Figure 7(d), ReaxFF correctly reproduces the qualitative trend that the C–O bond length decreases with increasing pressure. However, the predicted slope is smaller than that of the MP2 reference, indicating that the present force field is less sensitive to compression in this extreme pressure regime. The deviation from the MP2 values is about 0.15% at 0 GPa and increases gradually to about 0.79% at 20 GPa. Although this pressure range is far beyond typical industrial $scCO_2$ operating conditions, this comparison provides an additional benchmark for evaluating the structural response of the force field under extreme compression, which may be relevant to planetary/interior science or solid-state phase transition studies.

The radial distribution function (RDF) is a critical tool in structural analysis, offering detailed insights into the spatial arrangement of atoms or molecules within a material. For $CO_2$, the RDF describes how the density of $CO_2$ molecules varies as a function of distance from a reference molecule, providing essential information about the short-range and long-range order in the system. By calculating the RDF, we can identify the most probable distances between pairs of $CO_2$ molecules, which correspond to peaks in the RDF plot. In Figures 8(a-b), we calculated the RDF of $CO_2$ using ReaxFF for a liquid $CO_2$ system at 233 K (density = 1.08 g/cm$^3$) and a supercritical $CO_2$ ($scCO_2$) system at 315 K (density = 0.81 g/cm$^3$), and compared them with other force fields, such as Born-Oppenheimer molecular-dynamics (BOMD), Murthy, Singer, and McDonald (MSM), and TraPPE from existing literature.[87] We also performed simulations with the Cygan potential to compare RDF against ReaxFF. For both ReaxFF and the Cygan potential, simulations were performed with 108 $CO_2$ molecules (same as ref.[87]) under the NVT ensemble for 1 ns with a time step of 0.25 fs. The average RDF was then calculated by averaging RDFs over the last 500 ps to minimize noise. It was found that ReaxFF shows excellent agreement in predicting the RDF compared to other interatomic potentials. Such a comparison with published literature ensures that the new force field reproduces the known structural characteristics of $CO_2$, confirming its reliability and effectiveness in capturing the intricate details of molecular interactions and arrangements.

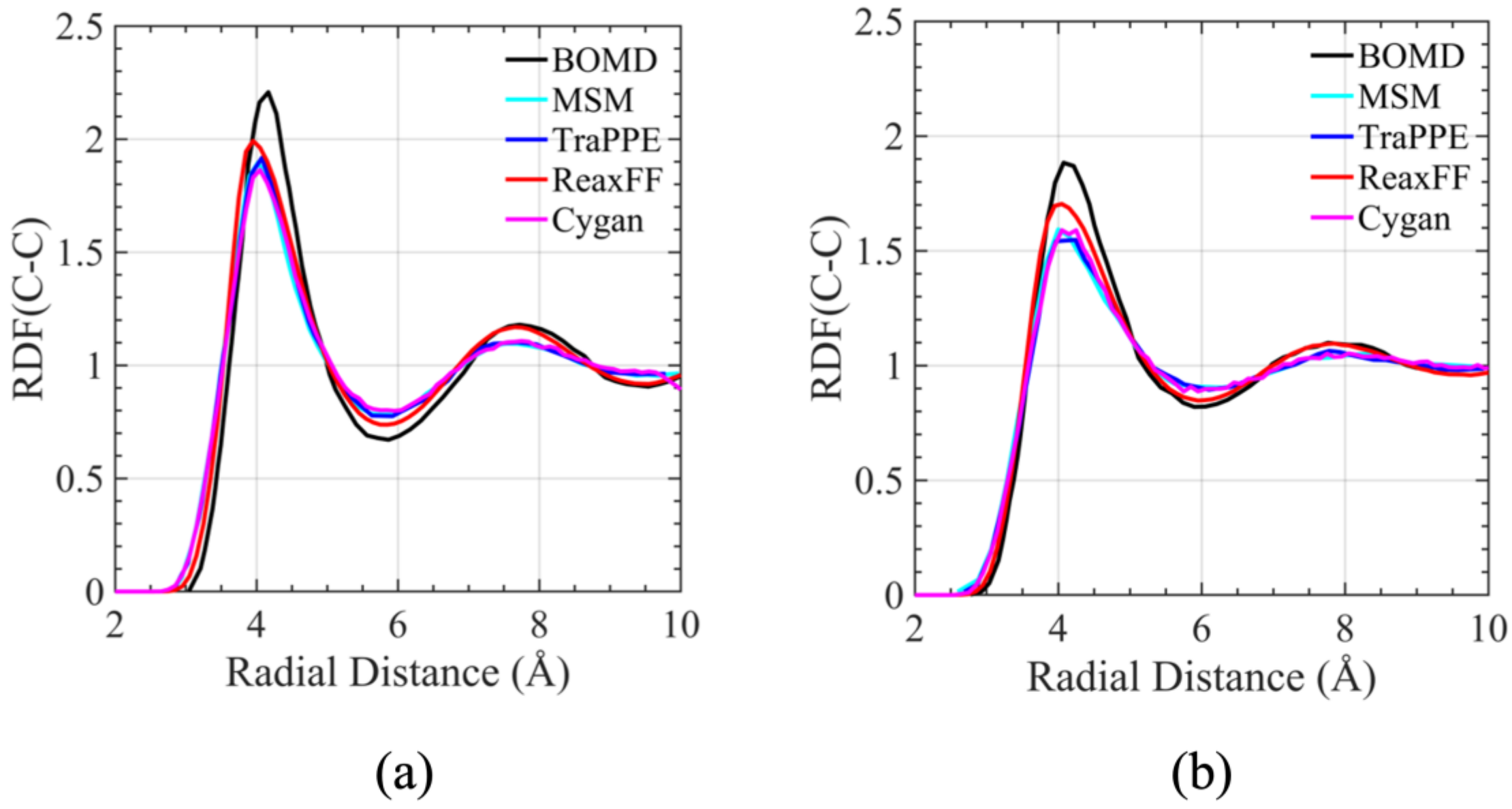

**Figure 8.** Comparison of $CO_2$ radial distribution function (C-C pair) between ReaxFF and other non-reactive force fields. (a) Radial distribution function plot for liquid $CO_2$ at 233 K (system density = 1.08 g/cm$^3$). (b) Radial distribution function plot for supercritical $CO_2$ at 315 K (system density = 0.81 g/cm$^3$).

### 3.3. ReaxFF Modeling of CO Oxidation in Supercritical $CO_2$

As mentioned earlier, in heavily backmixed CO-air systems and in high-temperature oxy-fuel or sc$CO_2$ flames, the O radical directly facilitates CO oxidation,[16,17,19–21] and one of our objectives is to investigate this phenomenon with our developed ReaxFF force field. It is worth noting that the production of $CO_2$ from CO and O is a highly exothermic reaction that releases energy in the form of heat. Consequently, freshly generated $CO_2$ can obtain a significant amount of kinetic energy (KE), distributed over translational, rotational and vibrational modes, as well as elevated potential energy (PE) due to the distortion of the molecular geometry away from the equilibrium linear structure. If the total energy (KE + PE) of the nascent $CO_2$ exceeds the bond dissociation energy (BDE) of $CO_2$ ($CO_2 \rightarrow CO + O$), which is approximately 125.73 kcal/mol (at 0 K),[88] $CO_2$ can dissociate back into CO + O. To assess the effect of finite temperature on this threshold, ReaxFF NVT simulations of isolated $CO_2$, CO, and O were performed at 330 K, yielding a thermally corrected BDE of approximately 122.7 kcal/mol, which is only ~3 kcal/mol lower than the 0 K reference and does not affect the conclusions of this analysis.

To investigate this phenomenon, a system comprising 20 CO molecules and 20 O atoms was considered within a 20 Å x 20 Å x 20 Å simulated box, yielding a system density of 0.18 g/cm$^3$. Subsequently, we conducted MD simulations within the microcanonical ensemble (NVE) for 1 ns, with a time step of 0.05 fs. It is noteworthy that systems containing radicals, such as O, need a reduced time step due to high local temperatures. For this system, it has been observed that when the time step was increased beyond 0.05 fs such as to 0.1 fs or 0.2 fs, the total energy was no longer conserved satisfactorily, particularly because of the very high local temperatures and steep forces associated with the radical reaction events. For this reason, a time step of 0.05 fs was adopted in order to maintain stable and physically reliable NVE dynamics. Figure S2 demonstrates that a time step of 0.05 fs successfully conserved the total energy during the NVE simulation of the 20 CO + 20 O system. We noted that the collision of CO and O resulted in the formation of numerous transient $CO_2$ molecules, which subsequently dissociated back into CO and O. Specifically, when $CO_2$ is created, if it is unable to disperse the substantial energy produced through collisions, one of the two oxygen atoms may detach from the $CO_2$ molecule. Figure 9(a) illustrates that the collision between a CO molecule and an O radical resulted in the formation of one $CO_2$ molecule. The oxygen originally bound to the carbon atom is designated as O(1), whereas the initial oxygen radical is designated as O(2). The freshly produced $CO_2$ could not achieve stability and ultimately dissociates again. However, the initially bound O(1) becomes separated this time. Figures 9(b) and 9(c) illustrate the variations in KE of the isolated O radical (O(2)) and initial CO(1) molecule prior to and subsequent to the impact, respectively. The CO(1) molecule collides with the O(2) radical at around 21.5 ps, resulting in the detachment of the O(1) atom from the newly formed $CO_2$ molecule at around 22.5 ps. Subsequent to the impact, the KE of the O(2) radical reduces by ~154 kcal/mol, whereas the KE of the CO(1) molecule increases by ~180 kcal/mol, surpassing the aforementioned BDE. Consequently, the CO(1) bond breaks. During the simulated period, $CO_2$ was continuously created and dissociated, resulting in a low level of stable $CO_2$ production. Therefore, it may be concluded that in a dilute environment, the conversion of CO and O into $CO_2$ is not efficient. It should be noted that the density of 0.18 g/cm$^3$ used in this dilute simulation is not representative of a true atmospheric-pressure gas-phase condition. Rather, this system serves as a reduced-density reference case relative to the dense $scCO_2$ environment (0.843 g/cm$^3$) studied subsequently. Simulating at truly atmospheric densities would require either impractically large simulation boxes or prohibitively long simulation times before sufficient reactive collision events could be observed within the nanosecond timescales accessible to ReaxFF MD, and is therefore outside the scope of the present study.

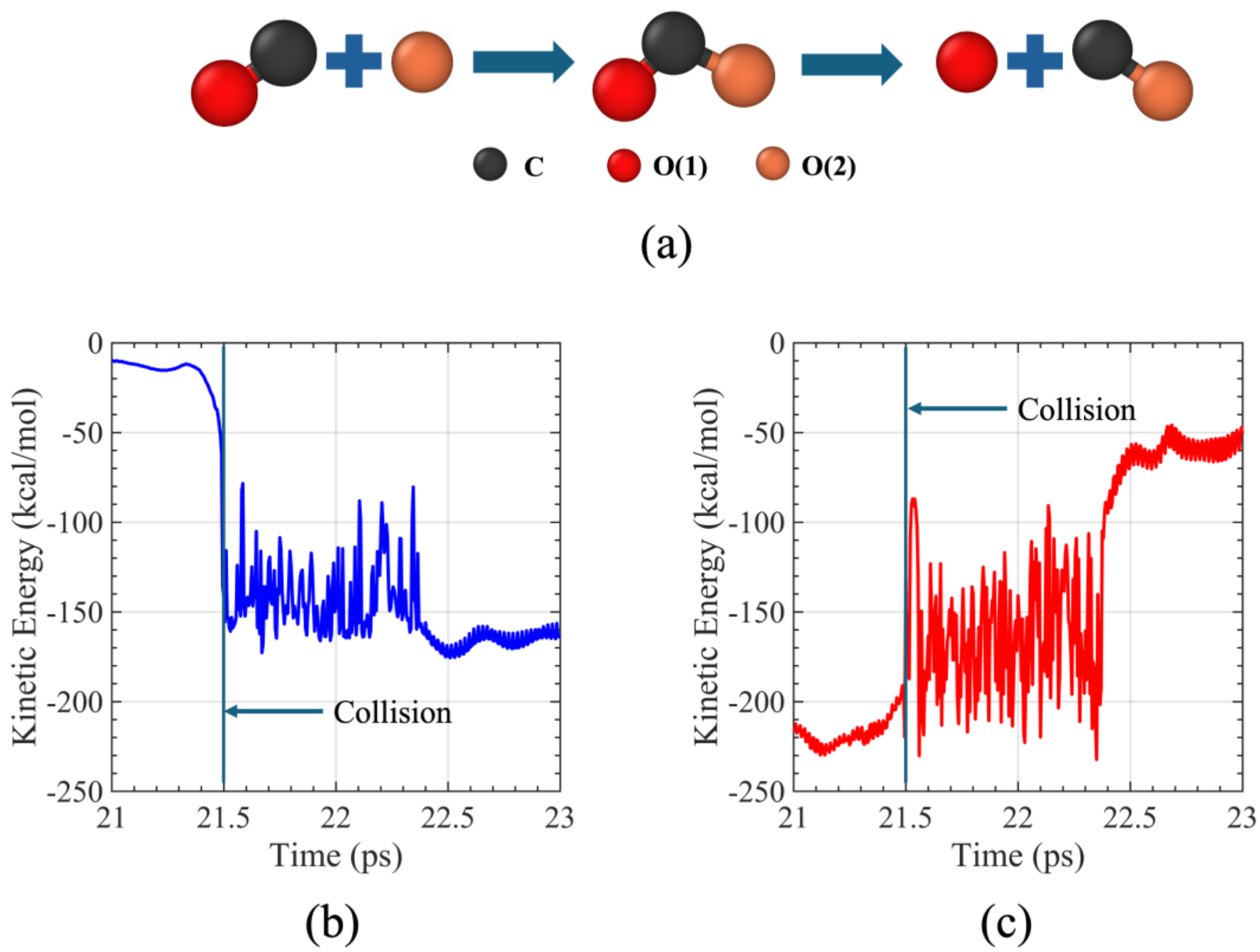

**Figure 9.** (a) Reaction of CO + O in a dilute system. The black sphere represents a C atom; the red sphere represents the O atom in the initial CO molecule (O(1)); the orange sphere represents an isolated O radical (O(2)). The newly generated $CO_2$ from this reaction did not sustain due to the large amount of KE generated due to the collision, and the initially bonded O(1) dislodged from the molecule. (b) represent the kinetic energy change of the O radical (O(2)) during the collision, and (c) represent kinetic energy change of the initial CO(1) molecule before and after the collision, respectively. The vertical lines in (b) and (c) indicate the start of the collision.

Consecutively, we investigated the CO + O reaction under an $scCO_2$ atmosphere. We initially utilized a system of 400 $CO_2$ molecules with a starting density of 0.843 g/cm$^3$ and conducted an NVT simulation at 330 K, employing a time step of 0.25 fs and a damping time of 100 fs. At equilibrium, the mean pressure of the system reached 25.3 MPa, and this pressure and temperature combination corresponds to the supercritical state of $CO_2$.[85] Afterward, 20 CO molecules and 20 O radicals were randomly added into the sc$CO_2$ system. We then simulated the reactants (CO + O) under the NVE ensemble, while regulating the solvent molecules ($CO_2$) at 330 K by a weak thermostat with a temperature damping constant of 1000 fs. The simulation was performed for 2 ns with a time step of 0.05 fs. Approximately 200 ps later, eight more stable $CO_2$ molecules were produced in the system, and after 2 ns, a total of eleven new $CO_2$ molecules were identified. The newly formed $CO_2$ molecules were observed to dissipate their excess total energy, comprising both elevated kinetic energy distributed over translational, rotational, and vibrational modes, and elevated potential energy arising from geometric distortion away from the equilibrium linear structure, through continuous collisions with the surrounding sc$CO_2$ matrix.

Figure 10(a) illustrates the simulation system comprising 20 CO molecules, 20 O radicals, and 400 sc$CO_2$ molecules used to investigate the CO + O reaction under supercritical conditions. Figures 10(b-d) show the kinetic energy, potential energy, and total energy profiles of a representative nascent $CO_2$ molecule from the moment of its formation to its fully stabilized state. Prior to the CO + O collision, the two reactant species carry a combined kinetic energy of only ~5 kcal/mol and a combined potential energy of approximately -180 to -190 kcal/mol, giving a total energy of approximately -185 kcal/mol. Immediately following the exothermic collision and $CO_2$ formation, the kinetic energy spikes dramatically to approximately 125-130 kcal/mol, reflecting the large amount of reaction energy released into atomic motion, while the potential energy drops sharply and fluctuates widely between -250 and -350 kcal/mol due to the intense structural distortion and vibrational excitation of the newly formed molecule. As a result, the average total energy of the nascent $CO_2$ immediately after formation is approximately -225 to -230 kcal/mol.

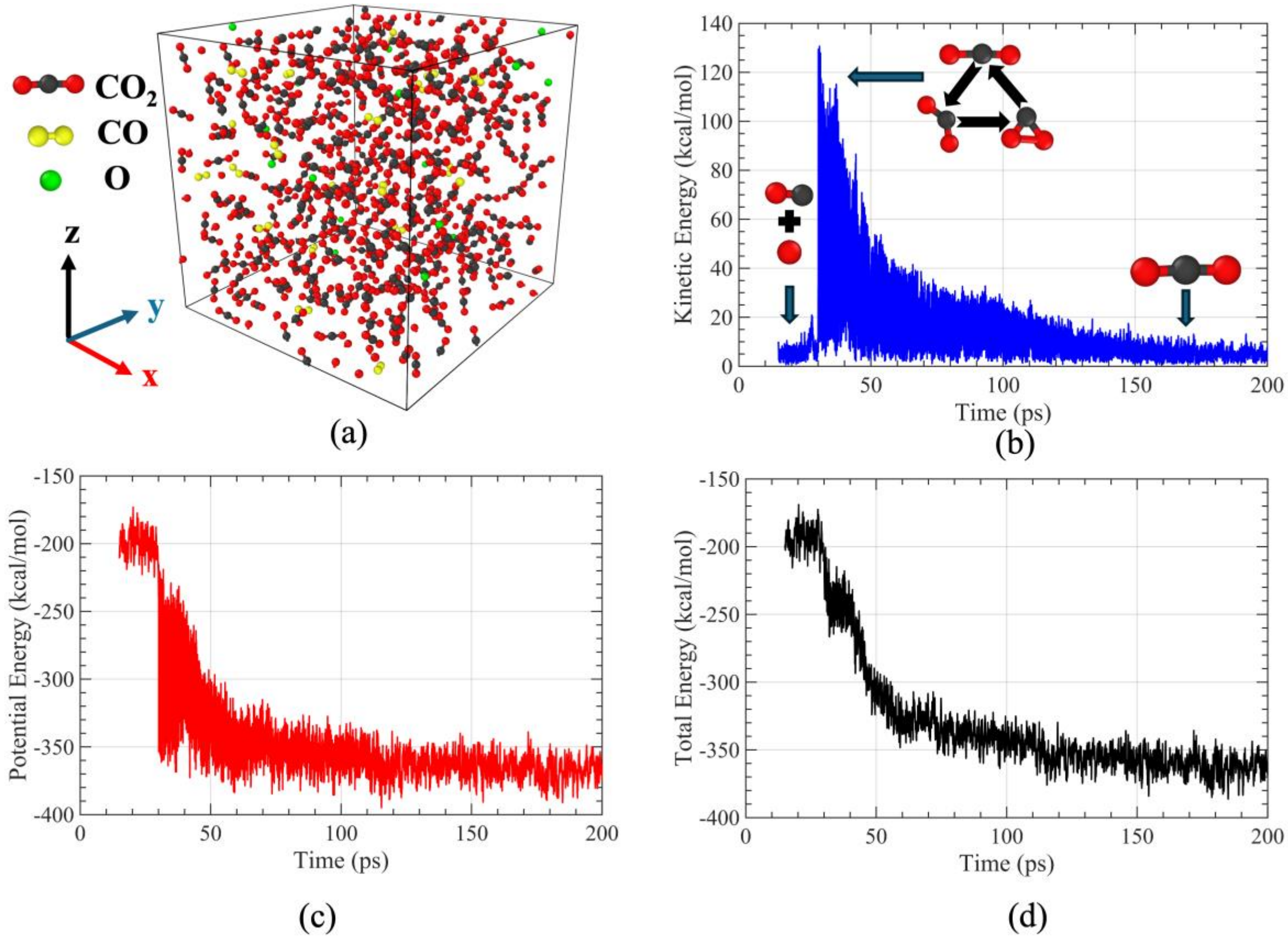

**Figure 10.** (a) Simulation system comprising 20 CO molecules, 20 O radicals, and 400 $scCO_2$ molecules. Simulation cell dimensions are 32.7 Å x 32.7 Å x 32.7 Å. (b) Kinetic energy, (c) potential energy, and (d) total energy profiles of a representative nascent $CO_2$ molecule from its formation through complete stabilization in the dense $scCO_2$ environment. Prior to ~30 ps, the energy profiles represent the combined energy of the separate CO and O reactants; from ~30 to 45 ps, the nascent $CO_2$ is formed in a highly energetic state with intense vibrational excitation and structural distortion, and subsequently loses energy continuously through collisions with the surrounding $scCO_2$ matrix, eventually reaching a stable, linear equilibrium geometry at approximately 150 ps.

Subsequently, through continuous collisions with the surrounding $scCO_2$ matrix, both the kinetic and potential energies gradually decrease until the kinetic energy returns to its thermally equilibrated value of ~5 kcal/mol and the potential energy settles near -360 to -365 kcal/mol, consistent with the ReaxFF-predicted minimized energy of a stable $CO_2$ molecule (~-366 kcal/mol). The KE and PE are dynamically coupled through vibrational motion and oscillate rapidly out of phase with each other throughout the stabilization process, as clearly visible in Figures 10(b) and 10(c). This gradual, multi-collision energy dissipation was observed consistently across all 11 nascent $CO_2$ molecules analyzed. Their initial total energies ranged from -220 to -233 kcal/mol, corresponding to an excess of 127 to 140 kcal/mol above the stable state, with an average energy dissipated of 133.9 ± 3.6 kcal/mol. The stabilization timescale ranged from 95 to 160 ps with an average of 112.4 ± 17.9 ps, confirming that the dense $scCO_2$ matrix acts as a sustained heat bath that progressively absorbs the excess total energy of the nascent $CO_2$ through repeated molecular collisions. The intense vibrational excitation of the nascent $CO_2$, manifested as large-amplitude bending and transient deviations from its linear equilibrium geometry, progressively diminishes as this excess energy is dissipated, and the molecule eventually reaches its stable, equilibrated configuration.

Figure 11 illustrates a representative individual collision event that contributed to the stabilization of a nascent $CO_2$ molecule. Figure 11(a) provides a snapshot highlighting the newly formed $CO_2$ (yellow) and its neighbors (green), defined as species located within a 4 Å distance from the nascent $CO_2$ at a specific time. The subsequent plots quantify the energy exchange during a 1.5 ps window. Figure 11(b) portrays a reduction of ~80 kcal/mol in the total energy of the new $CO_2$ molecule after a collision, whereas Figure 11(c) depicts a corresponding energy increase of ~70 kcal/mol in the neighboring molecules. The small discrepancy in energy change between the new $CO_2$ and its neighbors occurs due to the simultaneous energy exchange of these neighbors with their own surroundings. To provide a more mechanistic picture of this energy transfer, Figures 11(d) and 11(e) show the kinetic and potential energy changes of the nascent $CO_2$ during the same 1.5 ps collision window, confirming that both contributions decrease simultaneously during the collision event. Figure 11(f) further decomposes the kinetic energy into its translational component, computed as $KE_{trans} = 0.5 \times M \times |v_{cm}|^2$, and its combined internal (rotational + vibrational) component, obtained as $KE_{internal} = KE_{total} - KE_{trans}$. Here, M is the total molecular mass and $v_{cm}$ is the center-of-mass velocity. The results reveal that the translational KE accounts for only ~3.45 kcal/mol on average (~8% of the total KE), while the internal KE accounts for ~40.17 kcal/mol (~92% of the total KE). This demonstrates that the excess kinetic energy of the nascent $CO_2$ is predominantly stored in internal degrees of freedom rather than translational motion, consistent with the intense structural distortions observed in the trajectory visualizations. The small transient spike in translational KE observed around the collision event (~10-12 kcal/mol) represents a brief momentum transfer from the neighboring $scCO_2$ molecule that rapidly dissipates. Separating the rotational and vibrational contributions from one another would require projection onto instantaneous normal modes, which is not straightforward in a reactive MD framework due to continuous bond breaking and formation, and is identified as a direction for future work.

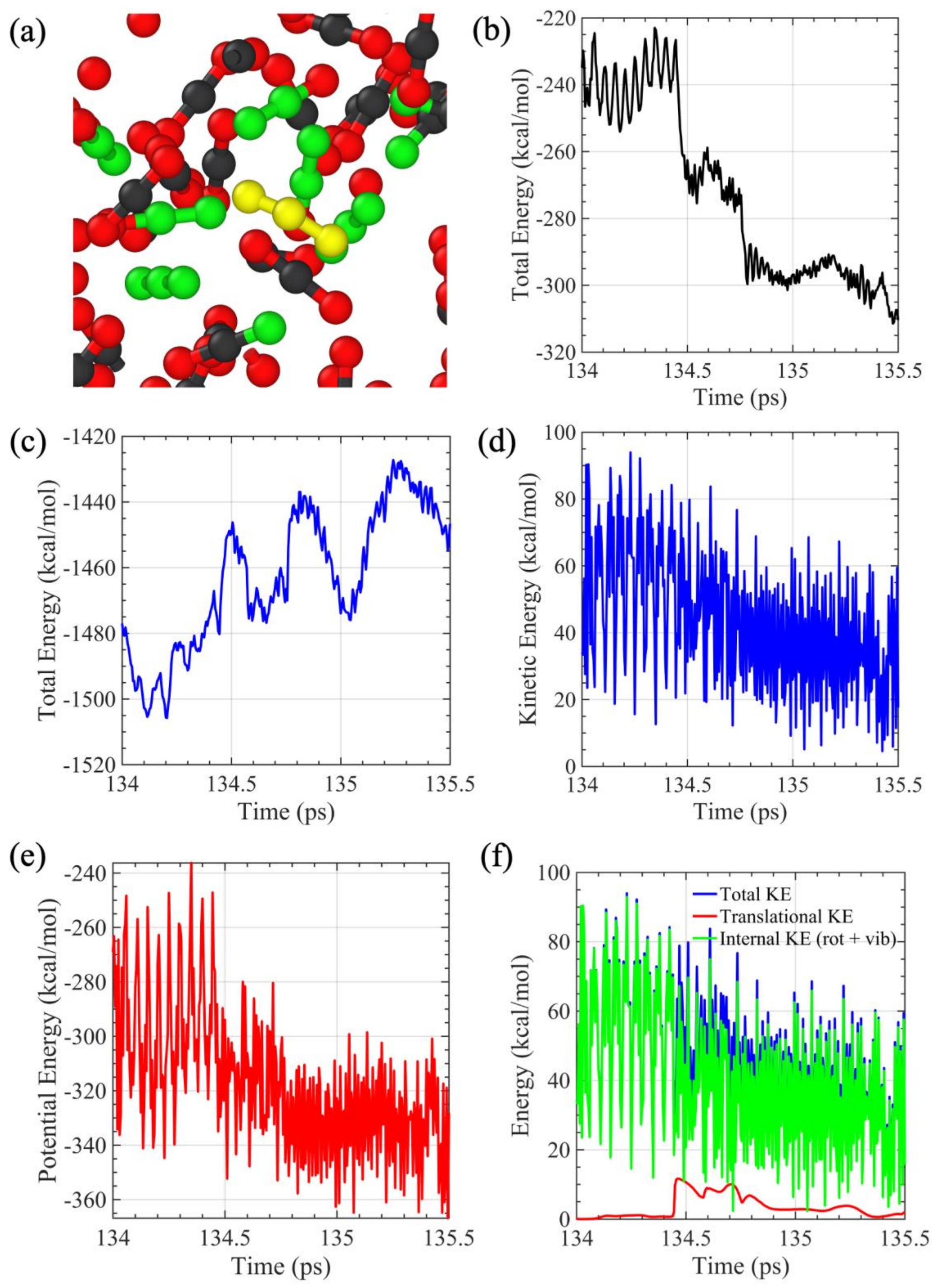

**Figure 11.** (a) Snapshot of a nascent $CO_2$ along with its neighbors. The new $CO_2$ is highlighted with yellow color, and its neighbors within 4 Å distances are highlighted in green. (b) Total energy

changes of a newly formed $CO_2$ molecule during a collision with neighbors. (c) Total energy changes of 10 neighboring molecules of the newly formed $CO_2$. (d) Kinetic energy and (e) potential energy of the nascent $CO_2$ during the same collision window. (f) Decomposition of the kinetic energy into translational (center-of-mass) and internal (rotational + vibrational) contributions, showing that the excess kinetic energy is predominantly stored in internal degrees of freedom.

This observation, in which the dense supercritical solvent aids in stabilizing highly energetic products from exothermic reactions by collisional energy transfer, is consistent with prior ReaxFF studies, including the research of Mirakhori *et al.*[60] on iodine recombination in supercritical xenon matrix. Our atomistic demonstration that newly formed $CO_2$ is stabilized via collisional energy transfer to the dense sc$CO_2$ environment yields important mechanistic insight for combustion kinetics and supercritical reactor design. It is worth mentioning that during the NVE simulations of the CO and O system in both dilute gas-phase and dense sc$CO_2$ environments, $CO_3$ was detected as a transient species. However, it was very short-lived, with a typical lifetime of around 20 fs, and then dissociated back into $CO_2$ + O. One possible explanation can be drawn from Figure 5(b), where both ReaxFF and DFT predict that the energy of $CO_3$ is higher than that of $CO_2$ + O. Therefore, the $CO_2$ + O state is energetically more favorable than $CO_3$, and the newly formed $CO_3$ cannot be stabilized under the present conditions. One such $CO_3$ formation and subsequent dissociation event is shown in Figure S3.

Note that chemical species were identified using ReaxFF bond-order-based species analysis with a bond-order cutoff of 0.3. The time evolution of species was monitored from the LAMMPS species.out file to detect reaction events. To distinguish transient reversible events from stabilized products, the corresponding trajectory frames were also examined directly in the visualizer. In the dilute system, newly formed $CO_2$ was repeatedly observed to dissociate back into CO and O, whereas in the $scCO_2$ environment, once formed, the new $CO_2$ molecules remained stable throughout the simulation window.

To examine the effect of system size on the predicted non-reactive properties of $scCO_2$, additional NVT simulations were performed for systems containing 500, 1000, 2500, 5000, and 11,000 $CO_2$ molecules. All systems were initialized at a density of 0.552 g/mL, corresponding to 20 MPa and 360 K according to NIST, and simulated for 1 ns. The equilibrium pressures obtained for these systems remained reasonably close to the target pressure (see Figure S4), indicating that the bulk thermophysical behavior of $scCO_2$ is not strongly affected by system size within this range, although smaller systems showed larger pressure fluctuations. Based on these results, the larger 11,000-molecule system was used for bulk equation-of-state validation (Figure 7), while smaller systems were employed in the training (Figure 6) and reactive simulations (Figure 10 and 11) to maintain computational efficiency and permit direct analysis of individual reaction events.

In practical $scCO_2$-based combustion and power cycle systems, the working fluid often contains trace amounts of impurities such as $H_2O$, $O_2$, and $N_2$, which may influence the third-body stabilization mechanism identified in this work. The present CO/O/$CO_2$ force field can in principle be extended to incorporate these species through merging with the CHON-2019 parameter set of Kowalik *et al.* [89], which provides an improved C/H/O/N description with particular focus on $N_2$ formation kinetics and radical interactions. Since CHON-2019 resides on the same ReaxFF combustion branch as the present force field, such merging is in principle straightforward, though test simulations at $scCO_2$ conditions and potential QM-based retraining would be required before deployment. From a physical standpoint, the three impurities are expected to influence the stabilization mechanism in qualitatively different ways. $N_2$ is a weak collider with low collisional energy transfer efficiency,[90] and its presence is not expected to meaningfully alter the dominant stabilization effect of the dense $CO_2$ matrix. $H_2O$, even at trace levels, is expected to be an efficient third-body collider due to its rich rovibrational mode structure and permanent dipole moment,[91] and may additionally participate in chemistry through hydrogen bonding and water-gas shift reactions. $O_2$ may introduce an additional CO oxidation channel via $CO + O_2 \rightarrow CO_2 + O$, and molecular simulations suggest that at trace concentrations it does not significantly perturb the bulk matrix structure.[92] A systematic atomistic investigation of these impurity effects using the extended force field represents an important direction for future work.

### 3.4. Comparison with Existing $CO_2$ Force Fields

Several classical force fields for $CO_2$ have been developed over the past four decades, each optimized for a specific class of properties. The early Murthy-Singer-McDonald (MSM)[93] model employed a three-site Lennard-Jones potential supplemented by a point quadrupole moment and demonstrated good agreement with solid-phase lattice vibrational frequencies and liquid-phase thermodynamics. The TraPPE[94] model of Potoff and Siepmann introduced optimized partial charges and van der Waals parameters to accurately reproduce the vapor-liquid coexistence curve and binary mixture phase equilibria of $CO_2$ with alkanes and nitrogen. Cygan et al.[81] subsequently extended earlier rigid three-site models by incorporating harmonic bond-stretch and angle-bend terms, enabling the prediction of vibrational spectra and diffusion coefficients in liquid and supercritical $CO_2$, as well as interfacial behavior with clay minerals. At a more accurate level, Born-Oppenheimer molecular dynamics[87] using DFT (PBE-D3) has been used to capture polarization effects, small deviations of the O-C-O angle from linearity (~174-176$^0$), and electronic absorption properties of liquid and $scCO_2$.

Despite their respective strengths, all of these models share a fundamental limitation: they are non-reactive. The MSM and TraPPE models employ rigid geometries in which C-O bond lengths and O-C-O angles are fixed, making bond dissociation physically impossible by construction. The flexible Cygan model employs harmonic potentials centered on the equilibrium geometry, which confines atoms near their equilibrium positions and similarly precludes bond formation or dissociation. BOMD, while electronically accurate, is computationally limited to systems of ~100 molecules over tens of picoseconds, and thereby, inadequate for simulating reactive events in dense sc$CO_2$ environments over nanosecond timescales and across thousands of atoms. None of these models can, therefore, address the central scientific question of this work: whether and how the dense sc$CO_2$ environment alters the fate of $CO_2$ formed from CO + O recombination.

The ReaxFF force field developed here overcomes these limitations through its bond-order formalism, in which interatomic interactions transition smoothly and continuously between bonded and non-bonded states as a function of atomic distances. Trained against MP2-level dimer energies, DFT-derived reaction pathways, and $CO_2$ equation-of-state data that span a wide density range (0.14–1.8 g/cm$^3$), the present force field simultaneously captures both the thermodynamic properties of bulk sc$CO_2$, validated against NIST pressure-temperature data and experimental radial distribution functions and the reactive chemistry of CO oxidation. This dual capability, reactive accuracy within a computationally efficient framework that scales to thousands of atoms and nanosecond timescales, represents the core advantage of the present model over all of its non-reactive predecessors and distinguishes it as a uniquely appropriate tool for studying third-body stabilization mechanisms in supercritical environments.

## Conclusions

In this study, we developed and utilized a new ReaxFF reactive force field to investigate the atomistic mechanism of the CO + O reaction in a supercritical $CO_2$ ($scCO_2$) environment. The force field parameters were trained using a comprehensive set of quantum mechanical (QM) data, supplemented by energy-density relationships obtained from the Cygan potential. The resulting model demonstrated good overall agreement with the reference values used for training. The force field was found to reproduce the pressure behavior of bulk $scCO_2$ with reasonable accuracy when compared to NIST data. In addition, the force field was validated against NIST pressure–density data at 360 K over the 10–20 MPa range, yielding an average deviation of ~5.62%, and against the pressure dependence of the C–O bond length in cubic $CO_2$-I up to 20 GPa, confirming that the correct qualitative trend of bond compression is captured even far beyond the intended application range. Furthermore, comparisons of the radial distribution function (RDF) for both liquid and $scCO_2$ showed strong accordance with other published non-reactive force fields, confirming the structural accuracy of ReaxFF. A dedicated comparison with existing $CO_2$ force fields, including MSM, TraPPE, Cygan, and BOMD, establishes that while each of these models captures specific structural or thermodynamic properties with high accuracy, none can address reactive events such as bond breaking and formation. The present ReaxFF force field uniquely bridges this gap by simultaneously reproducing bulk thermodynamic properties and reactive chemistry within a computationally tractable framework. Utilizing this novel force field, we observed that the formation of $CO_2$ from the CO + O reaction in a dilute system is highly inefficient. The high exothermicity of the reaction results in significant kinetic and potential energy transfer to the newly formed $CO_2$ molecule, which leads to its rapid dissociation without a third body to absorb the energy. During the NVE simulations, $CO_3$ was also detected as a transient intermediate in both the dilute and $scCO_2$ environments; however, it was very short-lived (~20 fs) and promptly dissociated back to $CO_2$ + O, consistent with the ReaxFF and DFT prediction that the $CO_2$ + O state is energetically more favorable than $CO_3$. However, when a similar reaction was modelled in a dense $scCO_2$ environment, the surrounding molecules functioned as an effective stabilizing medium. The newly formed, high-energy $CO_2$ product transferred its excess kinetic and potential energy through collisions with the solvent, leading to stability. This process was confirmed statistically across all 11 nascent $CO_2$ molecules identified during the 2 ns simulation: the average excess energy dissipated was 133.9 ± 3.6 kcal/mol, and the average stabilization timescale was 112.4 ± 17.9 ps. Analysis of the kinetic energy of the nascent $CO_2$ during representative collision events revealed that the excess kinetic energy is predominantly stored in internal (rotational and vibrational) degrees of freedom, accounting for ~92% of the total kinetic energy, while translational motion contributes only ~8%. This confirms that the $scCO_2$ matrix primarily quenches the intense vibrational and rotational excitation of the nascent $CO_2$ through repeated molecular collisions. The findings emphasize the active role of the supercritical $CO_2$ in stabilizing exothermic products and are anticipated to be valuable to the scientific community focused on combustion and high-pressure systems.

## Conflicts of interest

There are no conflicts to declare.

## Acknowledgments

The work was supported by the U.S. Department of Energy under contract No. DE-SC0022222. ACTvD and YKS acknowledge funding from the Office of Naval Research Grant through Grant

No. N00014-23-1-2725. MA would like to acknowledge the software and hardware resources of Stanford University used to perform QM computations. The ReaxFF MD simulations were performed on the Roar supercomputer managed by the Penn State Institute for Computational and Data Sciences.